\documentstyle[aps,prb,multicol,epsfig]{revtex} 

\begin{document}
\newcommand{\rhoS}{{{\mbox{\boldmath $\rho$}}}}

\title{  Effective action approach to strongly correlated fermion systems
}
\author{ R. Chitra$^\dagger$ and   Gabriel Kotliar}
\address
{Center for Materials Theory  and Department of Physics,
 Rutgers University, Piscataway, NJ 08854, USA.}

\date{\today}
\maketitle
\begin{abstract}
       We construct a new functional for the single particle Green's
function, which is a variant of the standard Baym Kadanoff functional.  
 The stability of the stationary 
 solutions to the new functional  is directly related to  aspects of the
 irreducible particle hole  interaction through the Bethe Salpeter equation.
 A startling aspect of this functional is that it allows a simple and rigorous
derivation of  both the
 standard and extended  dynamical mean field (DMFT) equations as stationary
conditions. Though the DMFT equations  were formerly obtained  only in the limit of infinite lattice coordination,
the new functional described in the work, presents a way of directly extending  
DMFT to finite dimensional systems, both on a lattice and in a continuum.
Instabilities of the stationary solution at the bifurcation point of the
functional,
signal the appearance of a zero mode at the Mott transition which then couples to physical quantities
resulting in divergences at the transition.
\end{abstract}
\maketitle
\vskip 1truecm 
\section{introduction}
 Functional methods have long been used as a tool to
study  interacting systems. These involve techniques
ranging from   simple weak coupling  perturbative methods to 
strong coupling expansions. 
Often, since it is not possible to sum the entire series of diagrams one
takes recourse to various approximation schemes in the hope that they
will capture some of the essential physics.
In addition, functional methods, have a long history of systematizing
these approximations.
A direct way of testing the validity of any approximation is through the effective action
approach to interacting systems.
Typically, the effective action is obtained 
 in terms of
some relevant variable, for e.g. the density, magnetization,  the order parameter,
the  single particle Greens function etc.,
 as a perturbative
series in the interaction.
A  well known  and frequently used functional is the 
Baym Kadanoff functional $\Gamma_{bk}$ (\ref{bk}) \cite{baym},  which is the effective action
 for the single particle
Green's function $G$. $\Gamma_{bk}$  gives  
the free energy of the system at its stationary point which are known to be
saddle points and not extrema.
In order to use variational approximations  on these functionals, 
the functional should  be convex. 
This is the case when one performs Legendre transformations with
respect to time or frequency independent quantities as 
in density functional theory \cite{fukuda}.
Very recently, a functional of the Green's function whose minimum
gives the local Green's function of the Hubbard model on the Bethe lattice
was investigated by one of us.
This functional has
several desirable properties, in particular it is 
a true extremum at the physical local Greens function.
\cite{gabi}. 
This  motivates us to search for generalizations
of the  functional of Ref.\onlinecite{gabi} to finite dimensions.

In this paper, we present a new
functional $\Gamma_{new}$ for the single particle Green's function,
for  which we can show that its  saddle point is in fact an  extremum,
in the case that the  irreducible vertex function in the particle hole
channel  has a definite sign (see below for a precise definition).
This functional has many other useful properties primarily in the context of
dynamical mean field methods which can  now be naturally viewed
as an approximation to this functional. The paper is organized as follows:
 first we introduce  the Baym Kadanoff
functional and our proposed functional
and obtain the  conditions
under which the saddle points of this functional are extrema.
We then  proceed to
show that the  dynamical mean field theory ( DMFT)
equations  are a simple outcome of
certain simple approximations on our functional,
hence providing some justification for direct extensions  of 
dynamical mean field theory to finite dimensions.
We also discuss some
potential applications  pertaining to  the Mott transition and the
problem of adapting tight binding basis for dynamical mean
field calculations.

\section{ The Baym Kadanoff functional}
\label{baymk}

 Consider  the partition function $Z$ (or equivalently the free energy $W$)
of
a system of electrons  moving  in a crystal potential   $V_c (x)$  with
Coulomb interactions  $V$  
\begin{eqnarray}
Z= \exp[-\beta W]&=& \int D[\psi \psi^\dagger] \exp    [-\int dx  \psi^{+}(x) [ \partial_{\tau}- {{\bigtriangledown^2} \over
{2m}} +V_{c}(x)]\psi(x) \nonumber \\
& - &
  \frac1{2} \int dx dx^{\prime} \psi^{+}(x)
 \psi^{+}(x^\prime) V(x-x^{\prime})
 \psi(x^\prime)
 \psi(x)]
\label{iham}
\end{eqnarray}
\noindent
Here   $x=({\bf x},\tau)$ denotes the space- imaginary 
time coordinates.
Our formulation is fairly general, it applies to both lattice models
or continuum models. In the lattice case, the space coordinate  ${\bf x}$
labels lattice sites and the integral symbol should
be interpreted as a summation symbol over the lattice sites.
We have omitted additional quantum numbers that can be carried by
the field $\psi$, such as spin  and  also orbital quantum numbers,
 in order not to clutter the notation.

  The effective action  for the single  Green's function $G$ i.e,
Baym Kadanoff functional, is defined as  
the Legendre transform of $W$  with respect to the interacting single particle
electron Green's function $G(x,y)$ i.e., 
\begin{equation}
\Gamma_{bk}[G]= W[J] + JG 
\label{bk1}
\end{equation}
\noindent
where, $W[J]$ denotes the free energy in the presence of an external
source field $J$ coupled to the electron Green's function.
Using the condition  $J=  {{\delta \Gamma_{bk}} / {\delta G}}$  to eliminate
$J$ in (\ref{bk1}), 
 we obtain  the Baym Kadanoff functional   
\begin{equation}
\Gamma_{bk} = [{\rm Tr} \log G
- {\rm Tr} [G_0^{-1}-G^{-1}]G + \Phi[G]]
\label{bk}
\end{equation}
\noindent
where, $G_0$ is the non-interacting Green's function 
 and $\Phi[G]$ is the sum of
all  two particle irreducible diagrams constructed using the interaction and
 $G$.
Stationarity yields 
 the  Dyson equation 
\begin{equation}
G^{-1} = G_0^{-1} - \Sigma[G]
\label{dyson}
\end{equation}
\noindent
where $\Sigma (x,y) = {{\delta \Phi} / {\delta G (y,x))}}$. 
At the saddle point (\ref{dyson}), $\Gamma_{bk}$ is just the
free energy of the system. An  effective action approach is most useful when
the  saddle points are extrema. This is  related  to   the
 stability of   stationary 
solutions,  which demands that  leading order fluctuations  around the stationary solution  (\ref{dyson}), increase the energy.
To obtain a formal criterion for stability, we  first
 consider the fluctuations around the stationary solution 
\begin{equation}
\delta \Gamma= \Gamma[G+ \delta G]  
- \Gamma[G] = \int\int \delta G(y,x) {{\delta^2 \Gamma} \over
{\delta G(x,y) \delta G(u,v)}} \delta G(v,u)
\label{bkstab}
\end{equation}
\noindent
where $\Gamma$ is any functional.
In the functional treatment, it is essential to note that
 any quantity $O(x,y)$ 
 is defined as
$O(x,y)= \langle x\vert O \vert y \rangle$ and hence  
should be viewed as a 
 matrix in $x$ and $y$.
 
To examine the stability of a saddle point of the functional
it is convenient to introduce a matrix ${\hat{ C}}$
 which acts on the Matsubara frequencies
 without affecting the spatial coordinates i.e.,
${\hat{C}} G(i\omega_n, {\bf x}, {\bf y}) = G(-i\omega_n, {\bf x} ,{\bf y}) $.
Using the  following property of Green's functions 
 $G(i\omega_n, {\bf x}, {\bf y})^{*} = G(-i\omega_n, {\bf x} ,{\bf y} $,
which we assume  to hold for the trial Green's functions as well,
we can rewrite 
(\ref{bkstab}) as
\begin{equation}
\delta \Gamma={ \sum_{n, m }} {\delta G(i\omega_n)
}^{*}   {{\hat {\chi}} (i\omega_n,i\omega_m) }{\delta G(i\omega_m)}
\label{form}
 \end{equation}
\noindent
The matrix  ${\hat \chi} \equiv  {\hat{ C}} \chi$ is defined by
$\chi=\delta^2 \Gamma/\delta G^2$.
While the frequency indices have been made explicit,
matrix multiplication in the real or momentum  space indices is implied.
For the saddle point to be an extremum (maximum or minimum) the form  
defined by (\ref{form}) has to be negative (positive).
For $\Gamma_{bk}$ 
\begin{equation}
{{\delta^2 \Gamma_{bk}} \over {\delta G(x,y) \delta G(u,v)}} = 
-G^{-1} (y,u)G^{-1} (v,x) +  \Gamma^{ph}(xy;uv)
\label{bkder}
\end{equation}
\noindent
Here $\Gamma^{ph}(xy;uv)=  \langle x y \vert \Gamma^{ph} \vert v u \rangle
={{\delta^2 \Phi} / { \delta G(x,y)\delta
G(u,v)}} $,  
 is the irreducible vertex which describes the irreducible
particle hole interaction in the two particle channel and should  be viewed as a tensor in the
indices  $x,y,u,v$. 
The analog of (\ref{bkder})  in momentum space is  
\begin{equation}
\chi_{bk}(p_1,p_2,p_3,p_4)=  - G^{-1} (p_2,p_3) G^{-1}(p_4,p_1) + \Gamma^{ph}(p_1,p_2,p_3,p_4) 
\label{bkmom}
\end{equation}
\noindent
  where,   $p=({\bf p},i\omega_p)$ with $\omega_p$ being the
Matsubara frequencies.
For $\Gamma=\Gamma_{bk}$, 
using (\ref{bkder}) in (\ref{bkstab}), we see that 
even in the non-interacting case, where $\Phi$ is zero, 
 the first term in (\ref{bkder}) makes $\Gamma_{bk}$ unbounded from below.
To further   illustrate this lack of stability in $\Gamma_{bk}$,   
 we consider the simple  example of an electron gas
with  Coulomb  interactions $V(x)= V({\bf x}) \delta(t)$.
Let $V({\bf q})$ be the Fourier transform of $V({\bf x})$.
In the  Hartree approximation,  $\Phi$ is  frequency independent and
 and rewriting the spin and momentum indices  explicitly,
\begin{equation}
\Gamma^{ph}({ p}_1,{ p}_2,{ p}_3,{ p}
_4) 
=\delta({ p}_1+ { p}_3- { p}_2-{ p}_4)V({\bf p}_3-{\bf p}_4)
\label{dhf}
\end{equation} 
\noindent
Substituting this in (\ref{bkmom})
\begin{equation}
\chi_{bk}({ p}_1,{ p}_2,{ p}_3,{ p
}_4
)= [-\delta({ p}_2-{ p}_3)\delta({ p}_1-{ p}_4)  G^{-1} ({ p}_2)
 G^{-1}({ p}_1) +   \delta({ p}_1+ { p}_3- { p}_2-{ p}_4)V({\bf p}_3-{\bf p}_4)] 
\label{chibkhf}
\end{equation}
\noindent
where,   $G^{-1}({ p}) = i\omega_p - \epsilon_p$ and  
 $\epsilon_p$ is the dispersion of the particles
shifted by the Hartree  self energy term $N V({\bf 0})$, with $N$ being
 the total number of particles. 

We see from (\ref{chibkhf})  that
 ${\hat\chi}_{bk}$, 
in the sector
of  zero total energy and momentum,
 ${\hat\chi}_{bk}$ 
 decouples into $2\times 2$  blocks in  the non interacting
limit,   with zero diagonal matrix elements
and  equal  off diagonal matrix elements   ${\omega_n}^2 + {\epsilon}^2 $.
Consequently, (\ref{form}) has positive and negative eigenvalues
even in the non interacting case.
The Hartree term adds the constant
$NV(0)$  to all entries of   ${\hat\chi}_{bk}$ .
This results in the stationary solution remaining a saddle point rather than
an extremum for small values of the interaction.
 Moreover,  (\ref{chibkhf}) shows that attractive interactions tend
to further destabilize  $\Gamma_{bk}$.

\section{A New  Functional}
\label{newf}
To 
 construct a  functional with a different
stability matrix,   
we modify $\Gamma_{bk}$ as follows:
\begin{equation}
\Gamma_{new}= \Gamma_{bk}  -Tr\log(1-JG) -TrJG
\label{newg}
\end{equation}
In simple terms, we have changed the value of the functional
away from the saddle point, by adding certain
source dependent terms which vanish quadratically as the source
is turned off.
Note that  though  $\Gamma_{new}$  is closely related to 
$\Gamma_{bk}$,
it is not a simple Legendre transform of the
free energy in the presence of a source coupled to a simple composite operator. 
Using  the relation  $J= \delta\Gamma_{bk}/\delta G= G^{-1} -G_0^{-1} +
  \delta\Phi/\delta G$ in (\ref{newg}) we get
\begin{equation}
\Gamma_{new}= -{\rm Tr} \log [G_0^{-1} - {{\delta \Phi} \over {\delta G}}]
-{\rm Tr} G {{\delta \Phi} \over {\delta G}} + \Phi
\label{new}
\end{equation}
\noindent
It is straightforward to check that 
 though  $\Gamma_{new}$ and $\Gamma_{bk}$  yield
 (\ref{dyson}) and  the same free energy
at stationarity, their higher order derivatives
evaluated at  stationarity  are different. This has profound implications
for the stability.
 
\begin{figure}
\centerline{\epsfig{file=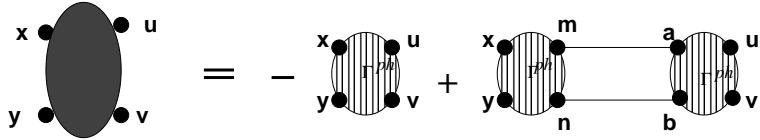,angle=360, width=10cm}}
\caption{ Diagrammatic representation of the (\ref{newder})
}
\label{var}
\end{figure}

The stability  of  a stationary
solution is  again described by  (\ref{bkstab}, \ref{form}), with $\Gamma$ replaced
by $\Gamma_{new}$.
In  the non-interacting
case, $\Gamma_{new}$ is trivially  
 a constant.
The second derivative of $\Gamma_{new}$   at stationarity  is 
\begin{equation}
{{\delta^2 \Gamma_{new}} \over {\delta G(x,y) \delta G(u,v)}} = 
- \Gamma^{ph}(xy;uv) +\int_{abmn} G(a,m) \Gamma^{ph}(mn;xy) G(n,b) \Gamma^{ph}(ba;uv)
\label{newder}
\end{equation}
\noindent
A diagrammatic representation of the above equation is shown in Fig.\ref{var}.
The integration over a variable $x$ denotes

\noindent
In momentum space, (\ref{newder}) reads 
\begin{equation}
{\chi}_{new} (l,m,p,q)= 
-\Gamma^{ph}(l,m,p,q) + \int_{k_1,k_2,k_3,k_4}  G(k_1,k_2)\Gamma^{ph}(k_2,k_3,l,m) G(k_3,k_4) \Gamma^{ph}(k_4,k_1,p,q) 
\label{mom}
\end{equation}
\noindent
where $\chi_{new}$ is the  Fourier transform of the second derivative of 
$\Gamma_{new}$.
In the translationally invariant case,  defining the momentum transfer
 ${ Q}= { l}-{ m}= { q}-{ p}$
(\ref{mom}) simplifies to  (Fig. 2)
\begin{equation}
\chi_{new}({ l},{ p};{ Q})=  
[-\Gamma^{ph}({ l},{ p};{ Q}) + \int_{{ k}}
  G
({ k}- \frac{  Q}{2})\Gamma^{ph }({ l},{ k};{ Q}) G({ k}+ \frac { Q}{2}) \Gamma^{ph}({ k},{ p};{ Q}) 
]
\label{bsf}
\end{equation}
\noindent

 For the stability analysis we limit ourselves to 
 consider  the case of zero transfer frequency
$\omega_Q=0$ since it is natural to expect that the leading
instability sets in first at zero frequency. This is known to be the case
of the two dimensional Hubbard model
within the full   parquet approximation 
as investigated  by Bickers \cite{bickers}.
To derive the conditions under which the stationary point is an extremum, 
we  analyze the form of  $\hat{\chi_{new}}$, and then 
derive general stability conditions for it.
The matrix $\Gamma^{ph}$
obeys the property ${\Gamma^{ph}}^\dagger = {\hat{ C}} {\Gamma^{ph}} {\hat{ C}}$
and so do  $(GG)$  and $\chi_{new}$, implying that
 ${\hat \chi_{new}}$ and ${\hat{ C}} {\Gamma^{ph}}$ are hermitian matrices.
The  particle hole kernel
which enters in the Bethe Salpeter equation can be rewritten as
$(1 - GG \Gamma^{ph}) = { \sum_i  {\lambda_i}  \vert\phi_i \rangle
 \langle\phi_i \vert}$  with
all the eigenvalues $\lambda_i $ positive as required by
the {\it stability}  requirements of the particle hole kernel. Moreover,
${\hat \chi_{new}}= - { \sum_i  {\lambda_i} C\Gamma^{ph} \vert\phi_i \rangle
 \langle\phi_i \vert}$. If $C\Gamma^{ph}$ is positive definite (negative definite), (\ref{form}) is
negative (positive) definite and hence the stationary solution denotes a maximum (minimum).
  Applying the same argument to $\Gamma_{bk}$, we note that the functional
is an extremum only if the Hermitian matrix $ CGG$ is positive or negative definite. Referring back
 to the previous section, we see that this is not
the case even in the non-interacting limit where this operator has eigenvalues
of both sign.To summarize,  the instability of the stationary solutions  of $\Gamma_{new}$ 
is linked  to  physical instabilities in the particle hole sector.
Since the eigenvalues of the Bethe Salpeter equation are always
positive when the solution is stable,  instabilities are signaled
by the  vanishing of an eigenvalue for the first time.

We now illustrate  the stability analysis of $\Gamma_{new}$ in the Hartree
approximation.
Using (\ref{dhf}) in (\ref{mom}), we see that
\begin{equation}
{\chi}_{new} ({ l},{ m},{ p},{ q})=
\delta({ l}-{ m}+{ p}-{ q})[ \int_{ k} G({ k}) G({ k}+{ l}-{ m}) V({\bf l}-{\bf m}) V({\bf m}-{\bf l}) - V({\bf l}-{\bf m})] 
\label{chinhf}
\end{equation}
\noindent

As before  we will consider  the case of zero  frequency and momentum transfer.
Summing over the internal frequencies
 $i\omega_k$ 
in  (\ref{chinhf})
we obtain
\begin{equation}
\chi_{new}({\bf l},{\bf p};{\bf Q})=V({\bf Q})
[-1 +V({\bf Q})\int_{\bf k} {{(n({\bf k}- \frac {\bf Q}{2}) -n({\bf k}+ \frac {\bf Q}{2}))
} \over {(\epsilon({\bf k}- \frac {\bf Q}{2}) -\epsilon({\bf k}+ \frac {\bf Q}{2
}))}}]
\label{chihf}
\end{equation}
\noindent
where, $n({\bf k})$ is the Fermi occupation function with the Hartree shifted
 chemical potential.

Taking the ${\bf Q} \to 0$ limit, 
\begin{equation}
\chi_{new} ({\bf l},{\bf p};{\bf 0})=- V({\bf 0})
[1 -V({\bf 0}) \int_{\bf k} {{\partial n({\bf k})} \over {\partial \epsilon_{\bf k}}}]
\end{equation}
\noindent
Note that $\chi_{new}$  viewed as a matrix in Matsubara frequencies
has constant  entries, and so does ${\hat\chi}_{new}$. All the eigenvalues
of this matrix are zero with the exception of one single eigenvalue proportional to 
\begin{equation}
-  V({\bf 0})[ 1+ V({\bf 0}) N({\tilde \mu})] 
\label{hfc}
\end{equation}
\noindent
 where $N({\tilde \mu})$ is the density of states corresponding
to the Hartree shifted chemical potential.
From (\ref{hfc}), we see that the Hartree approximation is again unstable
for repulsive interactions. However,  it is stable for attractive interactions
$V\to - V$, and becomes unstable when   
  $ { V}({\bf 0}) N({\tilde \mu})\to 1$ triggering 
 an instability in
the  particle-hole  sector. 
 This is the well known  phase separation
instability of fermions with attractive interactions. To investigate
superconducting instabilities one would have to introduce
anomalous Green's functions, and extend  the recent work of
Janis \cite{janis} who considered  the Baym Kadanoff functional,
to our functional. A more comprehensive discussion of  the stability of $\Gamma_{new}$ and the relative stability of
$\Gamma_{new}$ and $\Gamma_{bk}$ is left for future work.

\begin{figure}
\centerline{\epsfig{file=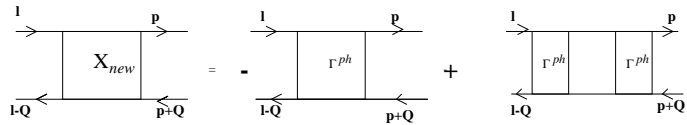,angle=360, width=9cm}}
\caption{ Diagrammatic representation of (\ref{bsf}) in momentum space
}
\label{bethe}
\end{figure}

\section{ Dynamical Mean Field Theory as an Approximation}
\label{sdmft}

We now focus on the various applications of the new functional introduced in
this paper. Despite the simple structure of both $\Gamma_{bk}$ and
$\Gamma_{new}$, it is often impractical and intractable to sum over the infinite
series of diagrams that contribute to $\Phi[G]$. Rather, various approximation
schemes are  used
to  obtain $\Gamma_{bk}$ ($\Gamma_{new}$)  and hence  $G$. The validity of these approximations
depend on whether (\ref{bkder}) (or (\ref{newder}) for $\Gamma_{new}$)  are positive or not. Typical  approximations
 range from   
 Hartree Fock and RPA  methods at weak coupling  to atomic limits for strong
coupling. Recent years have seen the development  of  
Dynamical mean field theory (DMFT) which is
 one of the very few
methods available to study  correlated electron systems in the weak, strong
and intermediate coupling regimes. Though this
method is exact only  in the limit of infinite dimensions, it has been
very successful in describing the physics of
realistic three dimensional systems, like 
the transition metal oxides.
DMFT  is formulated on a lattice and is  
 characterized by
self- consistency conditions (henceforth referred to as the
DMFT equations)  on the  full local Green's functions.
The form of these  
depend on the  non-interacting density of states of the lattice considered. 
These equations were derived using the cavity method or  local
perturbation theory in infinite dimensions \cite{rev}.
Consequently,  issues regarding
the fluctuations around the DMFT, 
extension to  finite dimensional lattices and systems in the continuum, 
 have been difficult to analyze.

Some of   the above issues can be resolved if DMFT  
 is  viewed as an approximation to $\Gamma_{new}$.
Here, for the first time, we show that the DMFT equations
can be derived directly from $\Gamma_{new}$, by
making the  ansatz 
 that $G$ is completely local i.e., $G({\bf r},{\bf r}^\prime,i\omega_n)=
G({\bf r},{\bf r},i\omega_n) \delta ({\bf r} -{\bf r}^\prime)$. 
Notice that our approach is fairly general. If we have in mind a lattice model
the delta function  $ \delta ({\bf r} -{\bf r}^\prime)$ should be understood
as a Kronecker delta function, while if we have a multiband model in mind
$ G_0 $ and $G$ should be viewed as matrices in the band and spin indices.
   Using the above ansatz,  the stationarity condition for $\Gamma_{new}$ yields
\begin{equation}
G({\bf r},{\bf r},i\omega_n)= [G_0^{-1} - { {\delta \Phi} \over {\delta G}}]^{-1}({\bf r},{\bf r})
\label{ndmf}
\end{equation}
\noindent
where the second term within the square brackets is a diagonal matrix with
entries
${ {\delta \Phi} / {\delta G}}={ {\delta \Phi} / {\delta G({\bf r},{\bf r},i\omega_n)}}$. 
The above equation should be viewed as a matrix in the spatial coordinates.
Interestingly,  (\ref{ndmf}) are exactly the   DMFT equations postulated
in Ref.\onlinecite{vlad} for non-translationally invariant systems.

The local ansatz applied to  $\Gamma_{bk}$ yields the
following equation
\begin{equation}
G^{-1}({\bf r},{\bf r},i\omega_n) =  G_0^{-1}({\bf r},{\bf r},i\omega_n)- {{\delta \Phi} \over {\delta G({\bf r},{\bf r})}} 
\label{bdmf}
\end{equation}
\noindent
The crucial point is that (\ref{bdmf}) and (\ref{ndmf}) become inequivalent  in  the local $G$ 
 approximation.
This  becomes transparent in momentum
space. 
Consider  a translationally invariant system
with  $G_0^{-1}= i\omega_n -\epsilon_k$  where $\epsilon_k$ is the
 energy dispersion of the non-interacting system.
Rewriting 
(\ref{ndmf})  in momentum space,  
\begin{equation}
G(i\omega_n)=  \int {{ \rho(\epsilon) d\epsilon} \over {( i\omega_n - \epsilon-  {{\delta \Phi} \over { \delta G(i\omega_n)}})}}
\label{dmft}
\end{equation}
\noindent
where, 
$G({\bf r}, {\bf r},i\omega_n)\equiv
G(i\omega_n)$ and 
 $\rho(\epsilon)$ is the  non-interacting density of states.
On the other hand, using the above form of $G_0^{-1}$ in (\ref{bdmf})  clearly
leads to unphysical conditions. 
Replacing  $\rho$ in (\ref{dmft}) by a lattice density of states in the limit
 of large dimensions yields the
 standard  DMFT equations.
The potential applications need to be explored further.

\section{Basis Optimization Criteria in DMFT Calculations}

Model Hamiltonians presuppose a choice of tight binding basis
and hence model parameters. In a previous paper \cite{previous}, we
argued that starting from the full Hamiltonian we can construct
a functional of the local Green's function which yields
the true local Green's function of the problem at stationarity. There,  DMFT viewed as
a theory for the local Green's function, is 
an exact theory but the functional which needs to be 
constructed is not known explicitly  (beyond a general diagrammatic
series discussed in ref \cite{previous}).
Here we adopt a different approach and would like
to view dynamical mean field theory as an
{\it approximation} to a simple functional.
When the saddle point fluctuations (\ref{form}) are positive,
 we can  gauge the quality of different approximations.

We first notice as in Ref.\onlinecite{previous} that 
different choices of tight binding orbitals
$\phi_\alpha$
where $\alpha$ denotes the band 
index, 
can be used to represent the exact one particle Green's function
\begin{equation}
G({\bf r},{\bf r}^\prime,i\omega_n) = \sum_{{\bf R}_1, {\bf R}_2} \phi_{\alpha}({\bf r}-{\bf R}_1) G_{\alpha \beta}
({\bf R}_1,{\bf R}_2,i\omega_n)
\phi_{\beta}({\bf r}^\prime-{\bf R}_2) 
\end{equation}

We can then insert the following approximate  trial Green's function into
our functional 
\begin{equation}
G({\bf r},{\bf r}^\prime,i\omega_n) = \sum_{{\bf R}_1, {\bf R}_2} \phi_{\alpha}({\bf r}-{\bf R}_1) G_{\alpha \beta}
({\bf R}_1,{\bf R}_2,i\omega_n)
\phi_{\beta}({\bf r}^\prime-{\bf R}_2) \delta ({\bf R}_1,{\bf R}_2)
\label{trial}
\end{equation}
\noindent
compute the trial value of  
  $\Gamma_{new}$ and optimize with respect to  
the $G_{\alpha \beta}$  first by solving the DMFT equations which
 expresses the  local lattice self
energy 
\begin{equation}
\Sigma_{\alpha\beta}({\bf R}_1, {\bf R}_1 , i\omega_n) =  {{\delta \Phi} \over {\delta G_{\alpha\beta}(
{\bf R}_1, {\bf R}_1 , i\omega_n)}} 
\end{equation}
\noindent
in terms of 
$  G_{\alpha\beta}(
{\bf R}_1, {\bf R}_1 , i\omega_n) $.
Another choice of orbitals  would lead to a different trial function
 (\ref{trial}),
so a functional perspective offers the possibility of optimizing
the basis of orbitals specifically for DMFT calculations.

\section{Derivation of Extended Dynamical Mean Field Equations}
\label{sedmft}

Ordinary dynamical mean field theory treats longer range interactions
at the level of the Hartree approximation.
An extended dynamical mean field approach has been
 suggested \cite{si} to partially remove this shortcoming.
 Recent work on the effects of long range Coulomb interactions within DMFT have shown that they
can substantially modify the behavior near a Mott transition \cite{mott},
changing the nature of the transition
from a continuous one  for short range interactions, to a discontinuous one.
In the light of these works and given the relevance of long range interactions, we present  a functional
derivation of  {\it extended}  DMFT equations
along the lines of Sec.\ref{sdmft}.
 
We
consider the system (\ref{iham})  for electrons with
some non-local interaction.                            
 Decoupling the  repulsive quartic interaction  term in (\ref{iham})  using a
Hubbard Stratanovich field $\phi$, we get

\begin{eqnarray}
\nonumber
Z&=& \int D[\psi \psi^\dagger \phi]  \exp    [-\int dx  \psi^{+}(x) [ \partial_{\tau}-
{{\bigtriangledown
^2} \over
{2m}} +V_{X}(x)]\psi(x) \\
& - &\frac1{2}\int dx dy  \phi(x)  \Pi_0^{-1}(x-y) \phi(y)   -  i\int dx \phi(x) \psi^\dagger(x) \psi(x)
\label{action}
\end{eqnarray}
\noindent 
If we chose $\Pi_0(x,y)=V(x-y)$  to be instantaneous in time,
then upon integrating out the the $\phi$
field we recover the interacting electron system of (\ref{iham}).
This decoupling can be extended to  attractive interactions  between the electrons by retaining
the same repulsive interaction between the phonons (so as to obtain a positive dispersion)
and   replacing
the factor $i$ in (\ref{action}) by unity. 

 Physically,  $\phi$ can be interpreted as a phonon field with
a particular
dispersion determined by $\Pi_0$. In fact, we can start with 
(\ref{action}) as a general action describing interacting electrons
and phonons, the limit of static Coulomb interactions being obtained
as a limiting case.
In momentum space, the term quadratic in the phononic field is $\phi( q) \Pi^{-1}_0(q) \phi(-q)$ with $\Pi_0 $ being the Fourier transform of the
interaction.
We now have a  system of interacting fermionic and bosonic fields.
Similar to Sec.\ref{baymk} , we calculate the free energy W[J,K] of the
electron-phonon
system, where   $J$ and $K$  are the source fields for the electron
Green's function $G$ and the phonon Green's function $\Pi$ respectively.
 Note that  in the present case, the phonon Green's function is nothing but
the full density density correlator or two particle Green's function of the
electrons.
The effective action functional i.e., the equivalent of the
 Baym Kadanoff functional for such a coupled system is now given by
the double Legendre  transform of $W[J,K]$         
\begin{equation}
\Gamma[G,\Pi]  = {\rm Tr} \log G
- {\rm Tr} [G_0^{-1}-G^{-1}]G -\frac1{2} {\rm Tr} \log \Pi
+\frac1{2} {\rm Tr} \Pi_0^{-1}\Pi +  \Phi[G,\Pi]
\label{bkex}
\end{equation}
\noindent
$\Phi$ is the sum of all  two particle irreducible  diagrams constructed
using the electron-phonon vertex.
The source fields are fixed by the following  conditions:
\begin{eqnarray} \nonumber
J&=&  G^{-1} -G_0^{-1}  + {{\delta \Phi} \over {\delta G}} \\
K&=&  \Pi^{-1} -\Pi_0^{-1}  - 2 {{\delta \Phi} \over {\delta \Pi}}
\label{sad}
\end{eqnarray}
\noindent
 The saddle point conditions on the above functional are given by
(\ref{sad}) with $J$ and $K$ set equal to zero.
The stability condition for this functional is given by
\begin{equation}
\int\int\sum_{i,j} \delta X_i(v,x) {{\delta^2 \Gamma} \over
{\delta X_i(x,y) \delta X_j(y,u)}} \delta X_j(u,v) \ge 0
\label{bkstabex}
\end{equation}
\noindent
where $X_i$ refer to $G$ and $\Pi$.                         
Following (\ref{newg})  we introduce a slightly
modified functional
\begin{equation}
\Gamma_{new}[G,\Pi]= \Gamma - Tr\log (1-JG)  -Tr JG   + \frac1{2}[ Tr\log (1-K\Pi)  +Tr K\Pi]
\label{func}
\end{equation}
\noindent
Substituting (\ref{sad}) in (\ref{func}), we get
\begin{equation}
\Gamma_{new}[G,\Pi]= -Tr \log\left (G_0^{-1} - {{\delta \Phi}
\over {\delta G}}\right) - Tr G {{\delta \Phi} \over {\delta G}}
 +{1 \over 2} Tr \log \left(\Pi_0^{-1} + 2 {{\delta \Phi} \over {\delta \Pi}}\right) - Tr \Pi {{\delta \Phi} \over {\delta \Pi}}  + \Phi[G,\Pi]
\label{newex}
\end{equation}
\noindent
These extra terms   tend to stabilize the functional by adding positive terms to
$\Gamma$.
The saddle point equations of (\ref{newex}) are
\begin{eqnarray}
{{\delta \Gamma_{new}} \over {\delta G}}&=& \left[ {1 \over {G_0^{-1} - {{\delta \Phi} \over
 {\delta G}}}}- G \right] {{\delta^2 \Phi} \over { \delta G\delta G}} + \left[{1 \over {\Pi_0^{-1}
+2  {{\delta \Phi} \over {\delta
 \Pi}}}}-\Pi \right] {{\delta^2 \Phi} \over { \delta G\delta \Pi}}  \\
{{\delta \Gamma_{new}} \over {\delta \Pi}}&=& \left[ {1 \over
{G_0^{-1} - {{\delta \Phi} \over
 {\delta G}}}}-G \right] {{\delta^2 \Phi} \over { \delta \Pi\delta G}} + \left[{1 \over
 {\Pi_0^{-1} + 2 {{\delta \Phi} \over {\delta
 \Pi }}}}-\Pi \right] {{\delta^2 \Phi} \over { \delta \Pi\delta \Pi}}
\label{sadex}
\end{eqnarray}

The saddle point solutions are obtained by equating the above
equations to zero.
The relevant solution is
$G^{-1} = G_0^{-1} - {
{\delta \Phi} / {\delta
 G}}$ and $\Pi^{-1}=\Pi_0^{-1} + 2{
{\delta \Phi} / {\delta
 D}}$ since  (\ref{func}) demands that when  $J=K=0$,
${{\delta  \Gamma_{new}} / {\delta X} }=
{{\delta  \Gamma} / {\delta X} }$.        
The  second order derivatives are given by
\begin{eqnarray}
{{\delta^2 \Gamma_{new}} \over {\delta G \delta G}}& =&
 \left[G {{\delta^2 \Phi} \over {\delta G \delta G}} G{{\delta^2
\Phi} \over {\delta G \delta G}}-{{\delta^2 \Phi} \over {\delta G
 \delta G}}\right] -2  \Pi {{\delta^2 \Phi} \over {\delta G \delta \Pi
}}   \Pi  {{\delta^2 \Phi} \over {\delta  G \delta  \Pi}}
 \\
{{\delta^2 \Gamma_{new}} \over {\delta \Pi \delta \Pi}}& =&- \left[2\Pi  {{\delta^2 \Phi} \over {\delta \Pi  \delta \Pi }}
\Pi {{\delta^2
\Phi} \over {\delta \Pi \delta \Pi}}+{{\delta^2 \Phi} \over {\delta \Pi
 \delta \Pi}}\right] + G {{\delta^2 \Phi} \over {\delta G \delta \Pi
}}  G  {{\delta^2 \Phi} \over {\delta G \delta G}}
 \\
{{\delta^2 \Gamma_{new}} \over {\delta G \delta \Pi}}& =&
G {{\delta^2 \Phi} \over {\delta G \delta \Pi}} G{{\delta^2 \Phi} \over {\delta G \delta G}}
-2  \Pi {{\delta^2 \Phi} \over {\delta \Pi \delta \Pi
}} \Pi  {{\delta^2 \Phi} \over {\delta G \delta \Pi}}  -{{\delta^2 \Phi} \over {\delta G \delta \Pi}}   
\end{eqnarray}
\noindent                      
To obtain the extended mean field equations, we make the
ansatz:  $G({\bf r},{\bf r}^\prime,i\omega_n)= G({\bf r},{\bf r},i\omega_n) \delta({\bf r}-{\bf r}^\prime)$    and
  $\Pi({\bf r},{\bf r}^\prime,i\omega_n)= \Pi({\bf r},{\bf r},i\omega_n) \delta({\bf r}-{\bf r}^\prime)$.
Minimizing with respect to $G$ and $\Pi$ we get         
\begin{eqnarray}
G({\bf r},{\bf r},i\omega_n)& =&  [G_0^{-1} - { {\delta \Phi} \over {\delta G}}]^{-1}({\bf r},{\bf r}) \nonumber \\
\Pi({\bf r},{\bf r},i\omega_n)& =&  [\Pi_0^{-1} +2 { {\delta \Phi} \over {\delta \Pi}}]^{-1}({\bf r},{\bf r})
\label{edmf}
\end{eqnarray}
\noindent
As before, ${ {\delta \Phi} / {\delta G}}$ and ${ {\delta \Phi} / {\delta \Pi}}$
 are local self energies and should be viewed as diagonal matrices in the space coordinates.
In the translationally invariant case, the  above reduce to the following equations
\begin{eqnarray}
G(i\omega_n)& =&  \int_{\bf q}  {{d{\bf q}}  \over { G_0^{-1} ({\bf q},\omega_n) - {{\delta \Phi}/{
\delta G}}}} \\
\Pi(i\omega_n)& =&   \int_{\bf q} { {d{\bf q}}   \over { \Pi_0^{-1} ({\bf q}) +2  {{\delta \Phi}/{\delta \Pi}}}}
\end{eqnarray}
\noindent
These are precisely the extended dynamical mean field equations obtained in
Ref.\onlinecite{mott}.
We therefore,  see that all  the dynamical mean field theories can be
seen as consequences of local approximations made on the new functionals.
 An analysis of stability  similar to the one leading
to the   conditions on (\ref{form})  is left for future work.

\subsection{Mott transition}
\label{mtrans}

Recently, the Mott (metal-insulator) transition in the  infinite dimensional
Hubbard model \cite{rev} was  revisited
 from  a Landau Ginzburg perspective \cite{gabi}.
 The central idea consisted of the introduction of a Landau
functional whose saddle point gives the (local) Green's function
of the problem. The Mott transition is then viewed as bifurcations
of the stationary points of this functional as the parameters of the theory
are
varied.
This approach was used to
clarify various issues related to the  zero temperature Mott transition
 in infinite
dimensions,  and to study the finite temperature Mott transition
at half filling in the infinite dimensional Hubbard model \cite{mar}.

Here we suggest that the functional  $\Gamma_{new}$ discussed in this paper,
 be viewed as a  generalization of the Landau functional
discussed in Ref.\onlinecite{gabi} to finite dimensions, and use it
to extend some  of the lessons learnt in infinite dimensions
to finite dimensions.
In particular,
we elaborate on  a connection between the existence
of a Mott transition and the divergence of the charge
compressibility,  conjectured on the basis of
pioneering numerical work by  Furukawa  and Imada \cite{imada}.                They  \cite{imada} discovered that the approach to
the Mott transition in two dimensions as a function of doping
is characterized by a {\it divergent}  compressibility.
By varying the hopping integrals, they demonstrated that
this is a general phenomena characteristic of the approach
to the Mott transition.                      
Here we use the following argument \cite{gkunpublished}. The Mott
transition is  but a bifurcation of the stationary solutions
of $\Gamma_{new} $ i.e.  solutions
of ${\delta \Gamma_{new}}[G] / {\delta G} =0 $,
as a control parameter $\alpha$ (which could stand for
chemical potential , temperature or   U  ) is varied.
For a bifurcation to take place
the matrix
$\chi_{new}=  {{\delta^2 \Gamma_{new}} \over
{\delta G(x,y) \delta G(u,v)}} $ has to be non invertible since otherwise
the equation:                   
\begin{equation}
{\delta G} =  -\left( {{\delta^2 \Gamma_{new}} \over {\delta G \delta G}}\right)_c^{-1}
\left({{\delta^2 \Gamma_{new}} \over {\delta G \delta \alpha}}\right)_c \delta \alpha
\label{delg}
\end{equation}
 \noindent
could be uniquely solved for $\delta G$.  In (\ref{delg}), the subscript $c$ denotes
the quantities evaluated at criticality. We have seen in 
Sec.\ref{newf} that when $\chi_{new} $ ceases to be  invertible,   an instability
sets in  the particle hole channel at zero momentum.
This implies a divergent compressibility unless there exist matrix elements
which cancel the contribution of the vanishing  eigenvalue of the Bethe Salpeter
kernel,
which signals the instability.
This argument  directly relates the existence of a Mott transition
to  the divergence of the compressibility supporting the
observation of Ref.\onlinecite{imada}.
We stress however that while our argument is fairly general
and very appealing in the light of the numerical results of
Ref.\onlinecite{imada}, it is based on assumptions which need
to be verified on a case by case basis.              
In fact, even in infinite dimensions,
the Landau functional at zero temperature has  some
very singular bifurcations
in its phase diagram   at which the compressibility does not
diverge.
This issue as well as the computation of matrix elements
in specific bifurcations
requires a problem specific  analysis  which at this
stage can  be performed rigorously only in the limit
of infinite dimensions\cite{gkunpublished}.

Finally it would be useful to extend the analysis  to spin dependent
Green's functions. This would  allow  us to understand the effects of
magnetism  in many  cases, examples being
 the  doping driven Mott transition in two dimensions,
where one approaches a magnetically ordered state  and the problem of
of the antiferromagnetic metal to antiferromagnetic insulator transition
in
infinite dimensions that we have studied recently
\cite{chitra1}.                                     

\section{Conclusion}
To summarize, we have constructed a new functional for the single particle
Green's function  and investigated its use as a  variational free energy.
We found that stability of the saddle point solutions was linked
to the Bethe-Salpeter equation in the particle hole sector.
{}From the perspective of applications,    the use of $\Gamma_{new}$
allows us to derive
DMFT and extended DMFT, directly in finite dimensions, for
non translationally invariant geometries.
Functional methods provide simple connections between one particle
Green's functions and particle hole instabilities, and we use these
ideas to discuss the compressibility near the Mott transition. 
Finally, we have shown that under some circumstances
it is possible  to understand whether the stationary points
of
$\Gamma_{new}$ are extrema or saddle points.
We feel that this is a significant advantage,
over the
Baym-Kadanoff functional $\Gamma_{bk}$,  when used to generate non 
perturbative approximations using trial Green functions.
There is a great deal of freedom, in adding sources
to $\Gamma_{bk}$ so as to convert its stationary points into extrema,
and constructions along the lines of our paper, but with different
additions of sources, so as to obtain more general stability conditions
than the ones we were able to obtain are well worth pursuing.
                                                              
\section{Acknowledgement}
This work was supported by NSF Grant 95-29138.

\end{document}